\documentclass{IEEEtaes}

\usepackage{color,array,amsthm}
\usepackage{graphicx}
\usepackage[caption=false, font=footnotesize]{subfig}
\usepackage{rotating}
\usepackage{pdflscape}
\usepackage{booktabs}
\usepackage{siunitx}
\usepackage{mathtools}
\usepackage{cuted}
\usepackage{multirow}
\usepackage{eqnarray}
\usepackage{amsmath}
\usepackage{placeins}
\usepackage{float}
\usepackage{lipsum}
\usepackage{flushend}
\usepackage{balance}
\usepackage{scalerel}
\usepackage{stfloats}
\usepackage{bigints}
\usepackage{textcomp}
\usepackage{tabularx}
\usepackage{cleveref}
\usepackage{booktabs}

\usepackage{makecell}
\usepackage{siunitx}
\usepackage{tcolorbox}

\usepackage{stackengine}

\usepackage[export]{adjustbox}

\usepackage{amsmath,amssymb,amsfonts,cite,graphicx,xcolor,color,multirow,array,eqnarray}
\usepackage{kantlipsum,lipsum,mathtools,tabularx,threeparttable,adjustbox,pdflscape,longtable,bigints}
\usepackage{flushend,booktabs,siunitx,textcomp,placeins,scalerel,cuted,breqn,csquotes,mwe,rotating,url}
\usepackage{algpseudocode}
\usepackage{algorithm}
\usepackage{setspace} 
\usepackage{nomencl}
\usepackage{etoolbox}
\makenomenclature

\usepackage{acronym}
\usepackage{array}

\begin{document}

\title{Multi-Orbiter Continuous Lunar Beaming} 

\author{BARIŞ DÖNMEZ}
\member{Graduate member, IEEE}
\affil{Polytechnique Montréal, Québec, CA} 

\author{YANNI JIWAN-MERCIER}
\affil{Polytechnique Montréal, Québec, CA} 

\author{SÉBASTIEN LORANGER}
\member{Member, IEEE}
\affil{Polytechnique Montréal, Québec, CA}

\author{GÜNEŞ KARABULUT KURT}
\member{Senior member, IEEE}
\affil{Polytechnique Montréal, Québec, CA}



\authoraddress{Barış Dönmez, Yanni Jiwan-Mercier, Sébastien Loranger, and Güneş Karabulut Kurt are the Poly-Grames Research Centre, Department of Electrical Engineering, Polytechnique Montréal and
Space Resources and Infrastructure Engineering Research Unit (ASTROLITH), Polytechnique Montréal, Quebec, H3T 0A3, Canada. 
(e-mail: \href{mailto:baris.donmez@polymtl.ca}{baris.donmez@polymtl.ca}). 
}


\maketitle

\begin{abstract}
In this work, free-space optics-based continuous wireless power transmission between multiple low lunar orbit satellites and a solar panel on the lunar rover located at the lunar south pole are investigated based on the time-varying harvested power and overall system efficiency metrics. The performances are compared between a solar panel with the tracking ability and a fixed solar panel that induces \textit{the cosine effect} due to the time-dependent angle of incidence (AoI). In our work, the Systems Tool Kit (STK) high-precision orbit propagator, which calculates the ephemeris data precisely, is utilized. Interestingly, orbiter deployments in constellations change significantly during a Moon revolution; thus, short-duration simulations cannot be used straightforwardly. In our work, many satellite configurations are assessed to be able to find a Cislunar constellation that establishes a continuous line-of-sight (LoS) between the solar panel and at least a single LLO satellite. It is found that 40-satellite schemes enable the establishment of a continuous WPT system model. Besides, a satellite selection method (SSM) is introduced so that only the best LoS link among multiple simultaneous links from multiple satellites will be active for optimum efficiency. Our benchmark system of a 40-satellite quadruple orbit scheme is compared with 30-satellite and a single satellite schemes based on the average harvested powers and overall system efficiencies 27.3 days so the trade-off options can be assessed from the multiple Cislunar models. The outcomes show that the average system efficiencies of single, 30-satellite, and 40-satellite schemes are 2.84\%, 32.33\%, and 33.29\%, respectively, for the tracking panel and 0.97\%, 18.33\%, and 20.44\%, respectively, for the fixed solar panel case.       

\end{abstract}


\nomenclature{$I(r,z)$}{Laser irradiance as a function of $r$ and $z$} 
\nomenclature{$r$}{Radial distance from the centre of the beam} 
\nomenclature{$I_0$}{Maximum irradiance at the beam center} 
\nomenclature{$w_0$}{Beam waist} 
\nomenclature{$w(z)$}{Beam radius limited by $1/e^2$ at $z$}
\nomenclature{$\eta_{t}$}{Power conversion efficiency from electrical power to optical power} 
\nomenclature{$P_I$}{Total input (electrical) power of an LLO satellite transmitter} 
\nomenclature{$\lambda$}{Optical wavelength} 
\nomenclature{$\theta$}{Beam divergence angle}
\nomenclature{$d_l$}{Transmitting telescope lens diameter}
\nomenclature{$A'$}{Area of the circular solar array}
\nomenclature{$L_{m}$}{Laser misalignment error loss factor}
\nomenclature{$\eta_{r}$}{Power conversion efficiency from optical power to electrical power}
\nomenclature{$P_{R_T}[n]$}{Time-varying received optical power for tracking solar panel}
\nomenclature{$P_{H_T}[n]$}{Time-varying harvested electrical power for tracking solar panel}
\nomenclature{$\overline{{P_{H_T}}}$}{Average harvested electrical power for tracking solar panel over 27.3 days}
\nomenclature{$\zeta_{T}[n]$}{Time-varying overall system efficiency for tracking solar panel}
\nomenclature{$\overline{{\zeta_{T}}}$}{Average overall system efficiency for tracking solar panel over 27.3 days}
\nomenclature{${P_{R_F}}[n]$}{Time-varying received optical power for fixed solar panel}
\nomenclature{${P_{H_F}}[n]$}{Time-varying harvested electrical power for fixed solar panel}
\nomenclature{$\overline{{P_{H_F}}}$}{Average harvested electrical power for fixed solar panel over 27.3 days}
\nomenclature{$\zeta_{F}[n]$}{Time-varying overall system efficiency for fixed solar panel}
\nomenclature{$\overline{\zeta_{F}}$}{Average overall system efficiency for fixed solar panel over 27.3 days}
\nomenclature{$n$}{Time index}
\nomenclature{$N$}{Maximum time index}
\nomenclature{${z}[n]$}{Time-varying minimum LoS distance between the solar panel center and multiple available LLO satellite transmitters}
\nomenclature{$\mathbf{Z}$}{Time step and satellite index dependent LoS distance matrix}
\nomenclature{$\psi[n]$}{Time-varying angle of incidence}
\nomenclature{$\mathbf{\Psi}$}{Time step and satellite index dependent angle of incidence matrix}
\nomenclature{$m$}{Satellite index}
\nomenclature{$M$}{Total number of simultaneously connected satellites to the lunar rover}
\nomenclature{$\mathbf{S}$}{Time-varying satellite selection matrix that contains ${z}[n]$ and corresponding $\psi[n]$ }
\nomenclature{$\hat{\textrm{\textbf{b}}}[n]$}{Time-varying transmitter boresight unit vector}
\nomenclature{$\hat{\textrm{\textbf{n}}}[n]$}{Time-varying solar panel surface normal unit vector}
\vspace{1 cm}
\printnomenclature[1.5 cm] 

\section{Introduction}
T{\scshape he} Moon, which is the natural satellite of the Earth, has great potential for many reasons. First, it enables not only Cislunar space missions but also beyond, such as deep-space missions (i.e., interplanetary), as it can be utilized as an intermediate destination \cite{coggins2024nasa, orbittypes, general1}. For instance, the maintenance of a spacecraft or recharge of a lunar rover can be realized on the Moon. Moreover, a non-governmental organization (NGO), Moon Village Association (MVA), envisions the settlement and exploration of the Moon by collaborating with more than 600 participants in academia, space agencies, government, and industry from 65 countries \cite{moonvillage}. Furthermore, the exploration of lunar regoliths in which precious minerals such as titanium and Helium-3 exist due to the asteroid impact has gained growing interest from space entities in different countries~\cite{lunarmining}. There are many smaller areas on the promising lunar south pole (LSP) region, such as the Amundsen, Malapert, and Shackleton areas \cite{LunarSouthPole}.

As there are many Cislunar orbits, their features need to be discussed and then compared in order to find the suitable one \cite{orbittypes}. Distant retrograde orbit (DRO) has a period of 14 days with a 70,000 km perilune, whereas it is oriented towards the Moon's equator. Earth-Moon Lagrange Point (EMLP)-2 halo orbits can have an amplitude of up to 60,000 km in addition to the 65,000 km distance between the lunar centroid and EMLP-2. Their period can last in the range of 8 to 14 days according to their lunar amplitude. As they are centered at EMLP-2, their orbiters can be utilized as a deep-space mission relay and/or lunar far side (LFS) relay \cite{donmez2024continuous}. Near-rectilinear orbits (NROs) are the subsets of EMLP-2 halo orbits, and perilune can take values between 2,000 to 75,000 km. Their period can be 6--8 days and are useful for polar coverage of the Moon. They are the Circular Restricted Three-Body Problem (CR3BP) as they are moving along with the Moon for keeping relatively stationary in the Earth-Moon plane. Elliptical lunar orbits (ELOs) have a period of roughly 14 hours as their perilune can be 100--10,000 km. However, their orientation heads toward the lunar equator. Frozen lunar orbits (FLOs) have a period of 13 hours with a perilune of 880--8,800 km. They can only be realized for specific inclinations, eccentricities, and energies, which make FLOs highly stable but, on the other hand, very challenging to construct. Prograde circular orbits (PCOs) have a period of 11 hours and approximately $75^\circ$ inclination in the Earth-Moon plane, whereas their perilune is larger than 3,000 km. Last but not least, low lunar orbits (LLOs) have a perilune of 100 km, with a period of 2 hours and without a constraint of inclination. These circular orbits are very flexible for Cislunar network design as their inclination can be $0^\circ$ for equatorial coverage or $90^\circ$ for LSP access. Besides, the LoS distance between a transceiver placed on the Moon and its orbiter is significantly short thus, the path loss can be minimized. Their major drawback is instability; hence, their revolutions do not repeat themselves unless there is a station-keeping \cite{orbittypes}.

Until now, the importance of lunar missions, promising locations on the Moon, and suitable Cislunar orbits have been discussed. Multiple wireless power transfer (WPT) technologies exist, such as the ubiquitous microwave and promising infrared laser. Microwave WPT provides higher power conversion efficiencies (PCEs) and coverages, especially useful for multi-point WPT, whereas growing distances cause significant path loss, and it can also interfere with the lunar devices working on the Moon \cite{donmezatp,gordon2023lasers}. On the other hand, free-space optics (FSO) technology which uses infrared laser transfers power in a collimated beam and hence the spot diameter on the receiver end becomes significantly shorter than its counterparts thus, the loss is mainly subject to the misalignment but an acquisition, tracking, and pointing (ATP) device can be used to mitigate the loss \cite{donmezatp}. Furthermore, the latter technology enables designers to implement smaller and lighter equipment in system design. Lunar dust \cite{naqbi2024opticalpowerbeaminglunar} is a challenge, especially for communication systems; however, the worst scenario happens after the receiver (e.g., a receiving telescope or a solar array) is covered with particles in post dust storms. Employment of an electric field or an ultrasonic vibration helps to clean the surface of a telescope lens \cite{gordon2023lasers}. 

WPT technologies were compared in terms of efficiencies and spot size diameters in \cite{marcinkowski2023lunar}. The scientists showed that microwave technology outperformed both FSO and millimeter-wave (mm-wave) in terms of input PCE. However, the spot diameters were computed for a transmitter diameter of 2.5 m with operational wavelengths of 1000 nm, 3.2 mm, and 6 cm at 100 km range as 0.098, 312, and 5856 meters, respectively. Since the receiver size is also a constraint, the highest portion of transmitted power can be collected by laser-based WPT systems at large distances when there is a perfect alignment. 

\vspace{-0.5 cm}
\subsection{Related Works}
\label{relatedworks}
There are multiple works on laser-based WPT in a lunar environment. In a survey study, the latest developments in laser-based WPT are discussed. The PCE of laser diodes are summarized according to their corresponding wavelengths ranging between 266 to 2100 nm. Moreover, current progress in optoelectric converters is listed based on their material, wavelength, PCE, and laser density (W/cm$^2$) \cite{zheng_wireless_2024}. The possibility of laser-based WPT from EMLP-1 and -2 to a lunar rover are calculated in \cite{Txaperturedia}. The array of laser diodes attached to the satellite provides WTP to a manned rover, and 30 kW of harvested power is the objective. Lasers with 10 W input power, 50\% PCE, and 800 nm wavelength are selected. On the other hand, Aluminium gallium arsenide (AlGaAs)-made solar cells with 50\% PCE and a pointing error of 0.025 $\mu$rad are considered. The circular solar array has a 4 m diameter, whereas the laser aperture size is 1 cm$^2$. In \cite{Nasa2000km3sat}, circularly orbiting equally separated three identical satellites that are 2000 km above the lunar surface provide WPT by laser to a lunar habitat and a rover for harvesting 1 MW and 75 kW, respectively. The transmitter aperture size of 8 m is considered to obtain a 1 m spot size. The Gallium Aluminum Arsenide (GaAlAs)-made solar cells with 32\% and 47\% of PCEs are used for the rover and habitat, respectively. A single laser with a wavelength of 850 nm in an array transmits 0.1 W of power with a diameter of 1 cm, whereas the laser array works at 70\% PCE. The solar array diameters are 7.5 and 7.6 m, and the pointing accuracy is 1 $\mu$rad. In \cite{bozek1995ground}, the lunar ground station supplies WPT to a spacecraft on the Moon. A laser diode with a wavelength of 800 nm, a PCE of 30\%, and a power of 1 W is considered, whereas 20\% of PCE for solar cells is used. In \cite{kerslake2008lunar}, the total WPT efficiencies of surface-to-surface WPT on the Moon's surface by using copper cable, radiofrequency (RF), and laser technologies are investigated. A circular laser planar array of 2 m diameter with 800 nm wavelength with input PCE of 50\% is taken into account. An AlGaAs-made circular solar array with a diameter of 1 m and a PCE of 50\% is considered. The target is to transfer 10 kW to a maximum distance of 1 km with a maximum misalignment error of 0.0286$^\circ$. Various transmitter aperture diameters are used up to 3.5 m, whereas the receiver aperture diameter is fixed to 3 m. As a result, copper cable outperforms the FSO; however, RF-based WPT using 5.8 GHz frequency has the least efficient technology among the three at any distance beyond 1 km.      

In \cite{lopez2023lunar}, lunar constellations that comprise up to ten satellites are investigated for various orbits, such as a lunar circular orbit with 600 km altitude and lunar frozen orbit with given Keplerian parameters for Shackleton Crater and a DRO for the lunar equator. Shackleton Crater in LSP and the lunar equator are the target areas in which power of 100 kW is harvested and then stored in the batteries. Orbital dynamics are modelled by using CR3BP. GaAs-made solar cells with a PCE of 68.9\% are used. Simulink and Simscape tools of Matlab are used to model the battery charging and discharging, and current, voltage, and state of charge are presented as a function of time up to three orbital periods. Furthermore, a power budget analysis is conducted for various receiver elevation angle constraints ranging from 50$^\circ$ to 80$^\circ$. The wavelength of 1064 nm, transmit power of 30 kW, and input PCE of 50\% are taken into account. In \cite{donmez2024continuous}, continuous power beaming to LFS from the EMLP-2 and halo orbits around this libration point is considered along with the stochastic misalignment error. Multi-orbiter cases are investigated for providing continuous surface coverage on LFS. A simplistic approach in which the effects of third-body gravitational forces are neglected is used in orbit design. In 
\cite{naqbi2024opticalpowerbeaminglunar}, the adverse effects of lunar dust on power beaming were investigated for the perfect alignment case. It is shown that the attenuation is significant for a ground-to-ground WPT case. In \cite{donmez2025hybridfsorflunar}, a multi-hop hybrid FSO-based and RF-based WPT is considered along with the random misalignment error. In the first hop, laser beaming is realized by SPS to the solar array on the LLO satellite. The harvested power is used entirely as the relay power. In the second hop, the RF power is transmitted to two separate locations, which are LSP and Malapert Mountain. The harvested powers at these locations are presented in stochastic models as random pointing error is considered in the first hop due to the major drawback of the FSO transmission.    

\begin{table}[!t]
  \sisetup{group-minimum-digits = 4}
  \centering
  \caption{Simulation Parameters of the Lunar Constellations}
  \label{tab:Kepler}
  \begin{tabular}{lllS[table-format=5]ll} 
    \toprule
    \toprule
     Semi-major axis & 1837.4 km  \\
     Eccentricity & $\approx0$   \\
     Inclination & $90^{\circ}$   \\
     Argument of perigee & $0^{\circ}$  \\
     Lunar rover location & $0^{\circ}\mathrm{E}\,,\,90^{\circ}\mathrm{S}$ \\
    \midrule 
    \textbf{True anomaly} \\
    30 satellites & $0^{\circ}:12^{\circ}:359^{\circ}$ \\ 
    40 satellites & $0^{\circ}:9^{\circ}:359^{\circ}$ \\
    \midrule 
    \textbf{Longitude of ascending node} \\
    Single orbit & $0^{\circ}$  \\
    Double orbits  & $0^{\circ}$, $90^{\circ}$  \\
    Triple orbits  & $0^{\circ}$, $120^{\circ}$, $240^{\circ}$  \\
    Quadruple orbits  & $0^{\circ}$, $90^{\circ}$, $225^{\circ}$, $315^{\circ}$  \\
    \bottomrule
    \bottomrule
\end{tabular}
\vspace{-0.2 cm}
\end{table} 

\subsection{Contributions}
\label{contributions}
It is very challenging to design a realistic orbital system as there are many force factors that need to be considered in astrodynamics. In our proposed continuous laser-based WPT model, we consider many aspects of orbital dynamics by utilizing System Tool Kit (STK) \cite{ansys_stk} for realistic and applicable scenarios. Then, we consider both tracking and non-tracking solar panels in our multi-orbiter Cislunar WPT analyses. 

Our contributions are as follows:
\begin{itemize}
\item In our Cislunar LLO WPT, STK high-precision orbit propagator (HPOP), which computes ephemeris data by using numerical integration of differential equations, is utilized for realistic orbiter movements. The gravitational force models of third bodies (e.g., Earth), solar radiation pressure (SRP), and Moon radiation pressure (e.g., albedo) are taken into consideration.

\item Many studies consider very limited simulation durations, such as a couple of periods or days; however, a full revolution of the Moon around the Earth that takes 27.3 days \cite{MoonEarthFactSheetNASA} is considered in our work. Interestingly, the initial deployment of the lunar constellations is significantly altered due to the aforementioned forces thus, short-duration simulation results cannot be used straightforwardly for longer projections, and hence, this makes our work realistic and applicable in real life.    

\item Many LLO satellites and orbits that provide continuous FSO-based WPT during a revolution of the Moon around Earth are assessed by using the STK HPOP to obtain precise data that will be useful in real-life space applications. It is found that 40 satellites enable the establishment of continuous WPT during 27.3 days.

\item Our continuous WPT system model enables simultaneous power transfer options by multiple orbiters. Thus, we introduce the satellite selection method (SSM) as presented in Algorithm \ref{alg:SatSel}, which aims to utilize only a single satellite which offers the highest system efficiency among many others at each time step. The results of this distance-based method are demonstrated in Fig. \ref{fig:3}. 

\item A solar array with tracking ability and a fixed solar array, which is affected adversely by the ever-changing angle of incidence (AoI), are compared in terms of harvested power and overall system efficiency throughout the Moon revolution around the Earth. It is found that the average harvested power decreases from 332.86 W to 204.43 W due to \textit{the cosine effect}. The overall system efficiencies of the single, 30, and 40 satellite cases as a function of time are presented in Fig. \ref{fig:5}.

\end{itemize}

\begin{figure*}[!t]
\centering
\subfloat[]{
	\label{subfig:1a}
	\includegraphics[clip, scale=0.26]{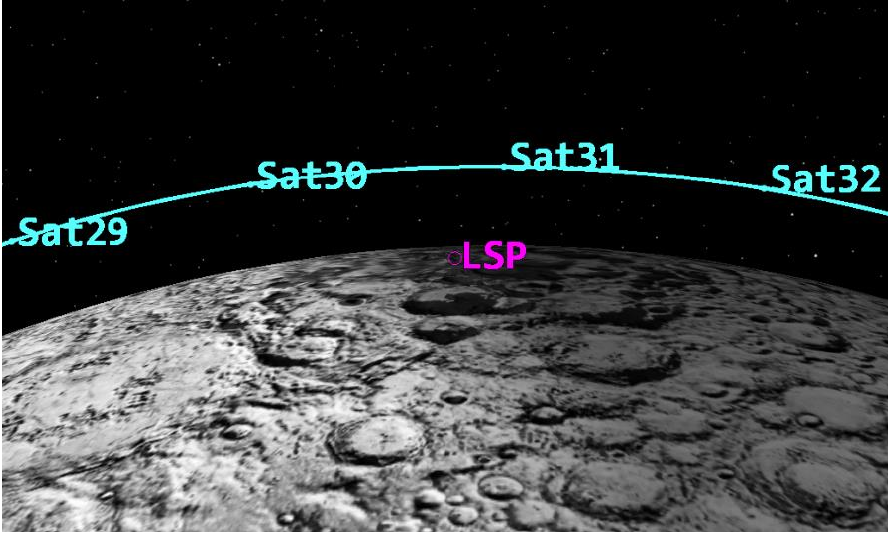}
	 }
\subfloat[]{
	\label{subfig:1b}
	\includegraphics[clip, scale=0.26]{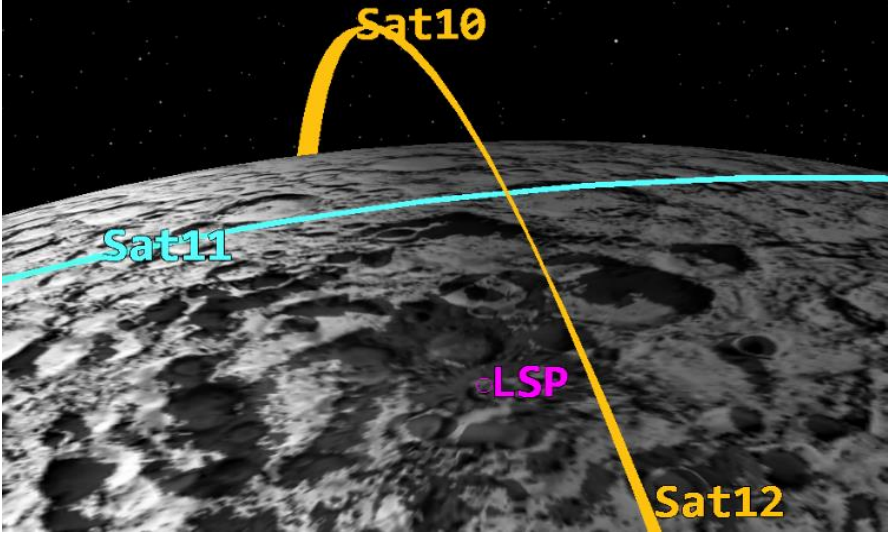}
	 }
\subfloat[]{
	\label{subfig:1c}
	\includegraphics[clip, scale=0.263]{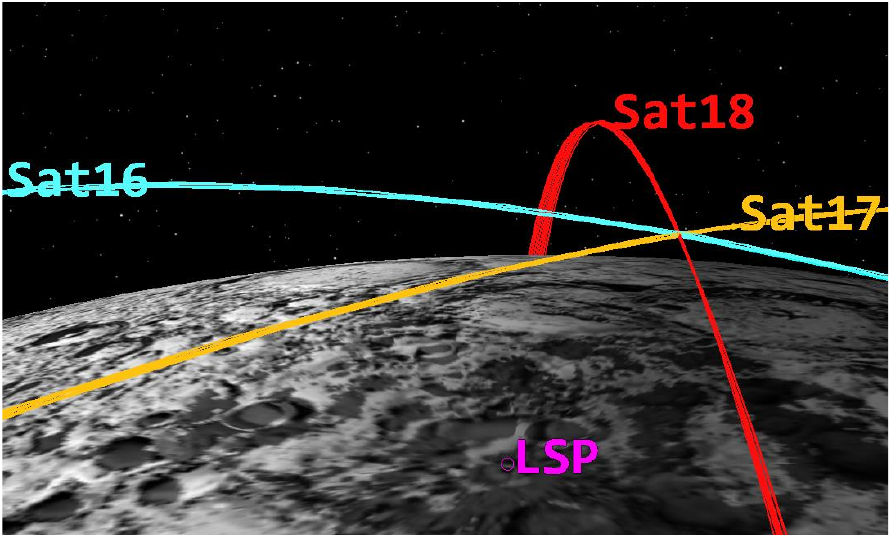}
	 }
\subfloat[]{
	\label{subfig:1d}
	\includegraphics[clip, scale=0.263]{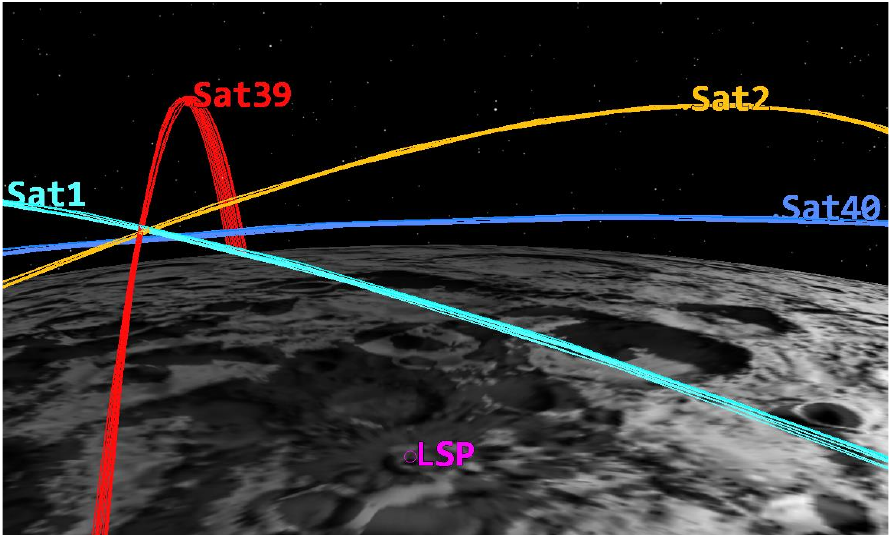}
	 }
\caption{Illustration examples of the 8 orbit configurations showing LLO satellites and LSP at which solar panel on the lunar rover is located (a) single orbit with 40 satellites, (b) double orbits with 30 satellites, (c) triple orbits with 30 satellites, (b) quadruple orbits with 40 satellites.}
\label{fig:1}
\end{figure*}

\begin{table}[!t]
\centering
\caption{Assessment of 30-Orbiter Configurations}
\label{tab:OrbitSelection30}
\resizebox{\columnwidth}{!}{%
\begin{tabular}{|c|cccc|}
\hline
\textbf{Total Orbiter} & \multicolumn{4}{c|}{\textbf{Orbit Configuration}} \\ \hline
30                        & \multicolumn{1}{c|}{\textbf{Single}} & \multicolumn{1}{c|}{\textbf{Double}} & \multicolumn{1}{c|}{\textbf{Triple}} & \textbf{Quadruple} \\ \hline
\textbf{Accessible Time Indices}     & \multicolumn{1}{c|}{39,327}          & \multicolumn{1}{c|}{38,972}          & \multicolumn{1}{c|}{39,249}          & 39,087             \\ \hline
\textbf{Access Rate (\%)} & \multicolumn{1}{c|}{99.92\%}       & \multicolumn{1}{c|}{99.01\%}       & \multicolumn{1}{c|}{99.72\%}       & 99.31\%          \\ \hline
\end{tabular}%
}
\end{table}

\vspace{-0.4 cm}
\subsection{Organization}
\label{organization}
The remainder of this paper is organized as follows. In Section II, eight different satellite configurations, which are single, double, triple, and quadruple orbits with 30 and 40 LLO satellites, are evaluated based on the continuity during a Moon revolution around the Earth. In Section III, the laser-based WPT model is presented for the solar panel with and without the tracking ability. The impact of the cosine effect is illustrated, and the satellite selection method is introduced. In Section IV, the harvested power and overall system efficiency performance metrics are computed first, and then the results are compared between our benchmark model, which is 40-satellite quadruple orbit, and 30-satellite and a single satellite schemes for the two types of solar panels. The average values of the metrics over a Moon revolution allow us to evaluate the trade-off options. Finally, our work is concluded in Section V.  


\label{Sec:intro}
\vspace{-0.5 cm}
\section{Assessment of Orbit Configurations}
In our proposed model, uninterrupted laser-based WPT to the promising LSP is targeted by utilizing multiple LLO satellites. However, selecting a suitable number of orbits and orbiters is crucial to achieving continuous WPT, as presented in Fig. \ref{fig:1}.    


Various orbital configurations, such as single, double, triple, and quadruple orbital scenarios with multiple orbits, are evaluated during the revolution of the Moon. It should be noted that the distances between the orbiters in each orbit are changing due to the aforementioned applied forces, and hence, considering 27.3 days instead of a single day enables our models to be applicable in a real Cislunar environment. 

First, we considered 30 satellites with a \textit{true anomaly} increment of $12^{\circ}$, which defines the starting position of a satellite in an orbit. Another Keplerian element of \textit{longitude of ascending node} is used to design new orbits; for instance, $0^{\circ}$, $120^{\circ}$, and $240^{\circ}$ are selected to create an equally spaced triple orbits scheme as exhibited in Table~\ref{tab:Kepler}.

\begin{table}[!t]
\centering
\caption{Assessment of 40-Orbiter Configurations}
\label{tab:OrbitSelection40}
\resizebox{\columnwidth}{!}{%
\begin{tabular}{|c|cccc|}
\hline
\textbf{Total Orbiter} & \multicolumn{4}{c|}{\textbf{Orbit Configuration}} \\ \hline
40                        & \multicolumn{1}{c|}{\textbf{Single}} & \multicolumn{1}{c|}{\textbf{Double}} & \multicolumn{1}{c|}{\textbf{Triple}} & \textbf{Quadruple} \\ \hline
\textbf{Accessible Time Indices}     & \multicolumn{1}{c|}{39,360}          & \multicolumn{1}{c|}{39,353}          & \multicolumn{1}{c|}{39,360}          & 39,360             \\ \hline
\textbf{Access Rate (\%)} & \multicolumn{1}{c|}{100\%}       & \multicolumn{1}{c|}{99.98\%}       & \multicolumn{1}{c|}{100\%}       & 100\%          \\ \hline
\end{tabular}%
}
\end{table}

According to Table \ref{tab:OrbitSelection30}, it can be seen that a total of 30 orbiters in various numbers of orbits cannot achieve a continuous WPT to the solar array at the LSP. For instance, a single orbit with 30 equally-spaced satellites initially can provide at least a single LoS link with the solar array for 99.92\% of the simulation duration, which yields $N$ maximum time index of 39,360 with a sampling time of 1~minute. 

After this result, the number of orbiters is increased to 40 to re-evaluate the continuity of single, double, triple, and quadruple orbitals as the previous ones fail. The step size of the true anomaly is reduced from $12^{\circ}$ to $9^{\circ}$, as shown in Table \ref{tab:Kepler}. On contrary to the 30 orbiter models, single, triple, and quadruple orbital configurations with 40 orbiters achieve continuous WTP during 27.3 days, as exhibited in Table~\ref{tab:OrbitSelection40}. It can be inferred that the double orbit scheme is outperformed by the others in both 30 and 40 satellite configurations. Thus, it can be concluded that the single, triple, and quadruple orbits with 40 satellites can be selected, according to Table~\ref{tab:OrbitSelection40}. Our choice is in favor of the quadruple orbits scheme to reduce the congestion in each orbit. It should be noted that the initial position order of the satellites changes due to the application of the aforementioned external forces during the time interval of 27.3 days, and each satellite on an orbit follows similar but different trajectories with substantial clearances in between them. 

The LLO satellites are identified based on their true anomaly values on Jan 1, 2025, at 00:00, and for each increasing true anomaly value, another orbit is considered according to the number of different angles of longitude of the ascending node. For instance, the true anomaly and longitude of the ascending node of Satellite 4 are $36^{\circ}$ and $0^{\circ}$, respectively, in the 30-satellite triple orbit scheme. On the other hand, the true anomaly and longitude of the ascending node of Satellite 4 are $27^{\circ}$ and $315^{\circ}$, respectively, in the 40-satellite quadruple orbit scheme.


\label{Sec:orbitconf}
\section{FSO-based WPT Model}
\begin{figure*}[!t]
\centering
\subfloat[]{
	\label{subfig:2a}
	\includegraphics[clip, scale=0.30]{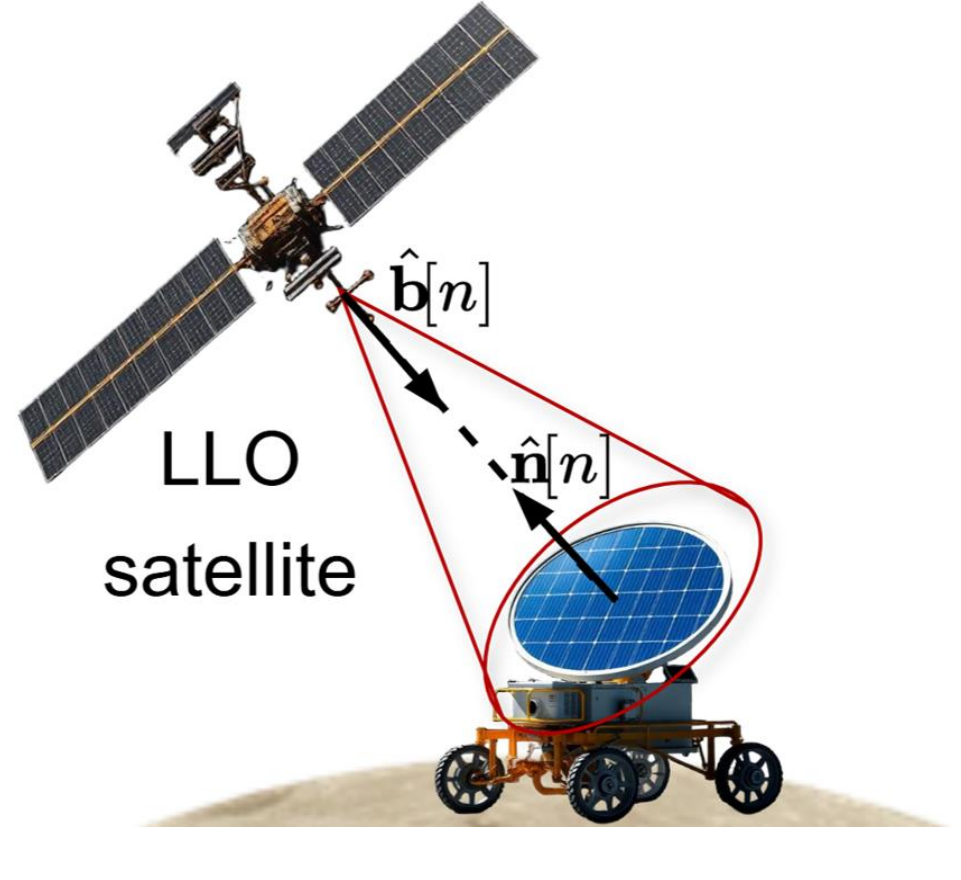}
	 }
\subfloat[]{
	\label{subfig:2b}
	\includegraphics[clip, scale=0.6]{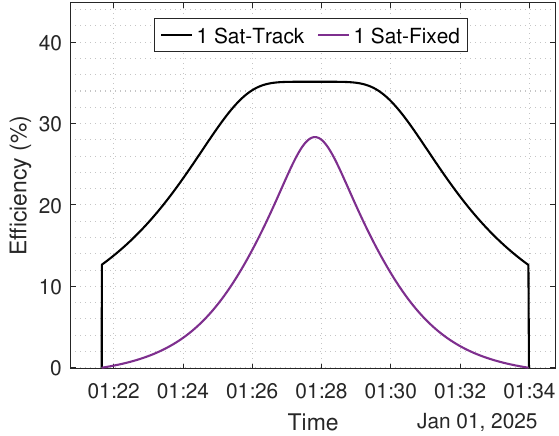}
	 }
     \subfloat[]{
	\label{subfig:2c}
	\includegraphics[clip, scale=0.32]{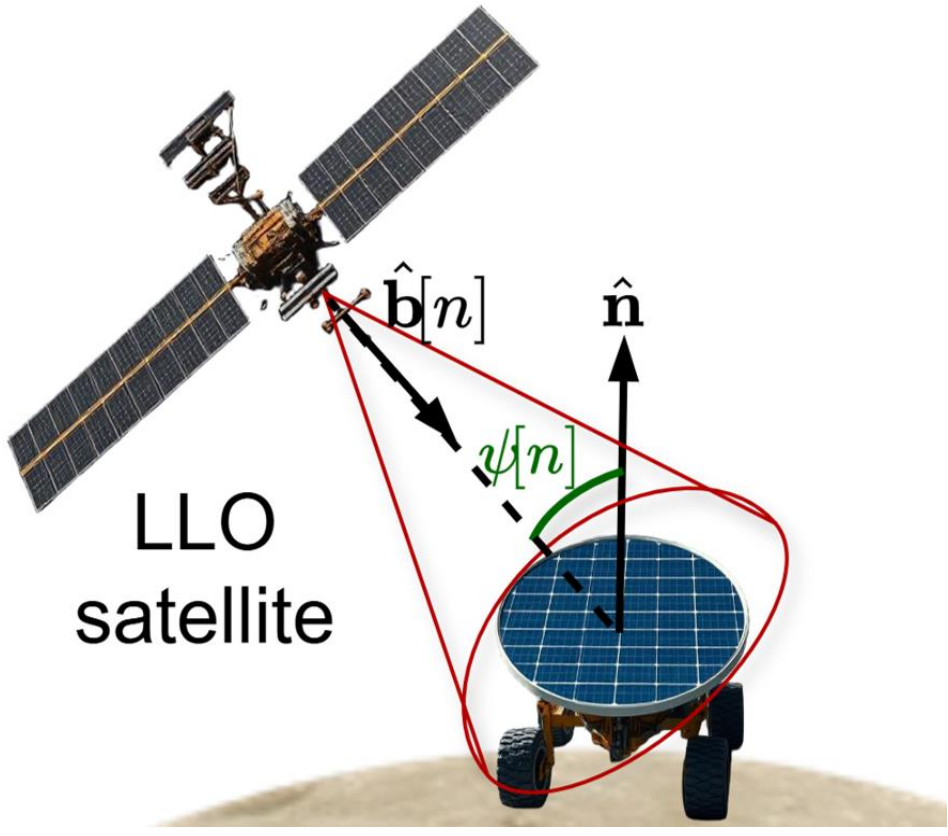}
	 }
\caption{Solar panel configurations and corresponding overall system efficiencies (a) tracking panel with aligned time-varying laser boresight unit vector {$\hat{\textrm{\textbf{b}}}[n]$} and solar panel normal unit vector {$\hat{\textrm{\textbf{n}}}[n]$}, (b) overall system efficiencies based on solar array features for a single satellite orbiting not directly above the rover, (c) fixed panel with fixed solar array surface normal unit vector {$\hat{\textrm{\textbf{n}}}$} and time-varying AoI $\psi[n]$.}
\label{fig:2}
\end{figure*}

The spatial distribution of the free-space laser beam irradiance at axial distance $z$ is modelled with a Gaussian wave model as follows \cite{majumdar2010free}:

{\small
\begin{equation}
{I}(r,z)={{I}_{0}}\exp \left( \frac{-2{\,{r}^{2}}}{w{{(z)}^{2}}} \right)=\frac{2\,{{\eta }_{t}}\,P_I}{\pi w{{(z)}^{2}}}\exp \left( \frac{-2{\,{r}^{2}}}{w{{(z)}^{2}}} \right), 
\label{EQ:1} 
\end{equation}
}%

\noindent where $r$, ${I}_{0}$, $w{(z)}$, $P_I$, and ${{\eta }_{t}}$ are radial distance from the z-axis, the maximum irradiance at the beam center, beam waist at distance z, total electrical transmit power, and electro-optical PCE respectively.

The $w{(z)}$ is defined as the beam radius that is bounded by the $1/\exp^2$ (13.5\%) of ${I}_{0}$ value at a distance $z$ as per the following equation \cite{majumdar2010free}.

{
\begin{equation}
w(z)={{w}_{0}}\sqrt{1+{{\left( \frac{z\lambda }{\pi {{w}_{0}}^{2}} \right)}^{2}}}, 
\label{EQ:2} 
\end{equation}
}%

\noindent where ${w}_{0}$ and $\lambda$ are the beam waist and the transmitter wavelength, respectively.

The beam waist ${w}_{0}$ can be found as follows:
{
\begin{equation}
{{w}_{0}}=\frac{\lambda }{\pi \theta }. 
\label{EQ:3} 
\end{equation}
}%

To provide a collimated beam with a very small $\theta$, larger ${w}_{0}$ and $d_l$ are required as can be inferred from Eq. (\ref{EQ:5})~and~(\ref{EQ:6}).

The laser beam divergence angle $\theta$ can be derived from a given laser telescope aperture diameter $d_l$ as \cite{OWCmatlab}
\begin{equation}
\theta\cong \lambda /{{d}_{l}}.
\label{EQ:4}
\end{equation}

\vspace{-0.5 cm}
\subsection{Tracking Solar Panel}
\label{Sec:sysmod:trackingsolarpanel}

In the tracking case, the time-varying panel surface unit vector {$\hat{\textrm{\textbf{n}}}[n]$} and laser boresight unit vector {$\hat{\textrm{\textbf{b}}}[n]$} maintain perfect alignment, as presented in Fig. \ref{subfig:2a}. In other words, the solar panel tilts itself in a way that incident laser rays become perpendicular to the collecting surface.

The transmitted power fraction that is collected by the solar array with an area $A'$ is time-dependent and can be found as \cite{farid2007outage}  
{
\begin{equation}
{{{P}_{R_T}}\![n]}=\int\limits_{A'}{\,I(r,{z}[n])\, dA}.
\label{EQ:5} 
\end{equation}
}%

As the solar panel is considered circular, the time-dependent collected portion of the transmitted power can be computed by
{\small
\begin{equation}
 {{{P}_{{R}_T}}\![n]}=\int\limits_{\phi'}{\int\limits_{r'}{\frac{2\,{{\eta }_{t}}\,P_I}{\pi w{{({z}[n])}^{2}}}\exp \left( \frac{-2\,{{r}^{2}}}{w{{({z}[n])}^{2}}} \right)\,r\,dr\,d\phi }}. 
 \label{EQ:6} 
 \end{equation}
 }%

The harvested power can be computed \cite{Shlomi_optimization_2004, ComprehensivePathLoss} as follows:
\begin{equation}
{{{P}_{H_T}}\![n]}={{P}_{R_T}}\![n]\,{{L}_{m}}{{\eta}_{r}},
\label{EQ:7}
\end{equation}
\noindent where ${{P}_{H_T}}[n]$ is the electrical harvested power and ${{\eta }_{r}}$ is defined as the solar cell PCE. In addition, ${{L}_{m}}$ is the misalignment loss factor, which takes a value between 0 and 1; thus, ${{L}_{m}}=1$ when there is a perfect alignment.

The average harvested power for tracking solar panel $\overline{{P_{H_T}}}$ can be calculated as follows:
\begin{equation}
{\overline{{P_{H_T}}}}=\frac{1}{N}\sum\limits_{n=1}^{N}{{{P}_{H_T}}\![n]}\ .
\label{EQ:8}
\end{equation}

The power efficiency ${{\zeta_T}\![n]}$ can be defined as the ratio of received and transmitted electrical power. 
\begin{equation}
{{\zeta_T}\![n]}=\frac{{{{P}_{H_T}}\![n]}\!}{P_I} \, \% \,.
\label{EQ:9}
\end{equation}

The average overall system efficiency for tracking solar panel $\overline{\zeta_{T}}$ can be found by
\begin{equation}
\overline{\zeta_{T}}=\frac{1}{N}\sum\limits_{n=1}^{N}{{{\zeta}_{T}}\![n]} \,.
\label{EQ:10}
\end{equation}

The quadruple orbits with 40 satellites, which enables continuous WPT during 27.3 days, are considered as our benchmark system model as mentioned in Section \ref{Sec:orbitconf}. However, it would be interesting to compare the performances of 30 satellites in quadruple orbits and a single satellite in a single orbit with our benchmark model. This will provide an opportunity to assess the trade-off options. The comparison between 30 and 40 satellites in quadruple orbits and a single satellite in terms of averaged values over a Moon revolution duration are exhibited in Table \ref{Table:Comparison} and discussed in Section \ref{Sec:results}. 

\vspace{-0.75 cm}
\subsection{Fixed Solar Panel}
\label{Sec:sysmod:fixedsolarpanel}

In the fixed case, the time-varying laser boresight unit vector {$\hat{\textrm{\textbf{b}}}[n]$} is not followed by the solar panel surface unit vector {$\hat{\textrm{\textbf{n}}}[n]$} as it maintains its alignment with the Moon surface normal unit vector, as presented in Fig. \ref{subfig:2c}. In other words, the solar panel keeps itself stationary, and the angle of incidence always becomes larger than zero; hence, a loss in the collected power appears as a result of the cosine effect.  
 
The received portion of the transmitted power is time-dependent and can be found as \cite{farid2007outage} 
{
\begin{equation}
{{{P}_{R_F}}\![n]}=\int\limits_{A'}{I(r,{z}[n])\,\cos(\psi[n])\, dA}, 
\label{EQ:11} 
\end{equation}
}%
\noindent where $\psi[n]$ is the angle of incidence which causes a degradation in received power. 

The collected portion of the transmitted power is computed for the circular solar panel as
{\small
\begin{equation}
{{{P}_{R_F}}\![n]}=\int\limits_{\phi'}{\int\limits_{r'}{\frac{2\,{{\eta }_{t}}\,P_I}{\pi w{{({z}[n])}^{2}}}\exp \left( \frac{-2\,{{r}^{2}}}{w{{({z}[n])}^{2}}} \right)\,\,\cos(\psi[n])\,r\,dr\,d\phi }}. 
\label{EQ:12} 
\end{equation}
}%

The harvested power for the fixed solar panel can be computed as follows:

\begin{equation}
{{{P}_{H_F}}\![n]}={{{P}_{R_F}}\![n]}{{L}_{m}}{{\eta}_{r}} \, .
\label{EQ:13}
\end{equation}

The average harvested power for the fixed solar panel $\overline{{P_{H_F}}}$ can be computed by

\begin{equation}
\overline{{P_{H_F}}}=\frac{1}{N}\sum\limits_{n=1}^{N}{{{P}_{{{H}_{F}}}}\![n]} \, .
\label{EQ:14}
\end{equation}

The power efficiency ${{\zeta_F}\![n]}$ for the non-tracking solar array can be obtained as follows:

\begin{equation}
{{\zeta_F}\![n]}= \frac{{{P_{H_F}}[n]}}{P_I} \, \% \, .
\label{EQ:15}
\end{equation}
 
The average overall system efficiency for the fixed solar panel $\overline{\zeta_{F}}$ can be calculated as follows:

\begin{equation}
\overline{\zeta_{F}}=\frac{1}{N}\sum\limits_{n=1}^{N}{{{\zeta}_{F}}\![n]} \, .
\label{EQ:16}
\end{equation}

In Fig. \ref{subfig:2b}, an overall system efficiency comparison between tracking and fixed solar panels is made for a single period of a single satellite. The single satellite (Satellite 1) orbit has a true anomaly of $0^{\circ}$ and a longitude of ascending node of $0^{\circ}$. The peak efficiencies are 35.14\% and 28.33\%, for the tracking and fixed panels, respectively. The reason for the discrepancy is that the orbiting satellite never passes as exactly perpendicular to the fixed solar array, or $\psi[n] \neq 0^\circ$, $\forall n$.

\begin{figure*}[!t]
\centering
\subfloat[]{
	\label{subfig:3a}
	\includegraphics[clip, scale=0.8]{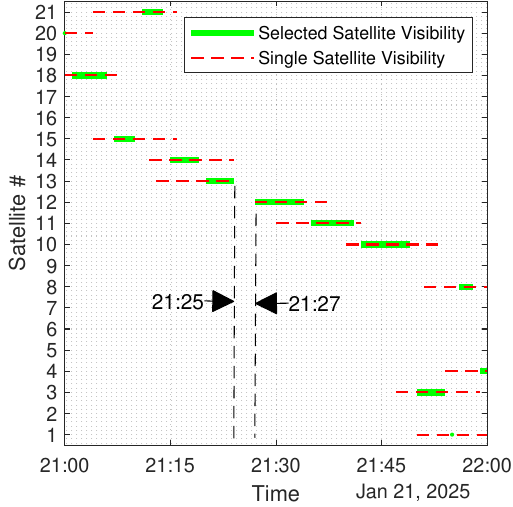}
	 }
\subfloat[]{
	\label{subfig:3b}
	\includegraphics[clip, scale=0.8]{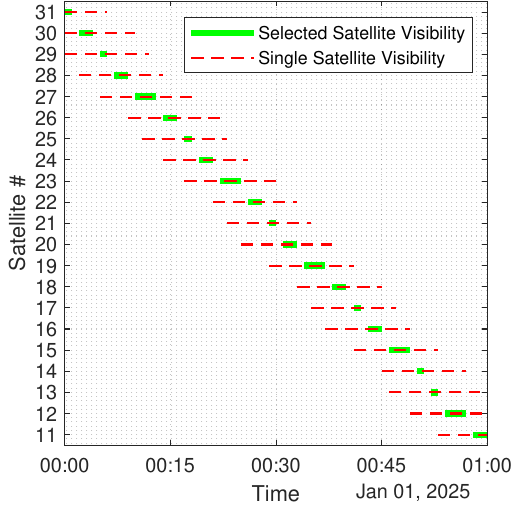}
	 }
\caption{Visibility intervals of quadruple orbit scheme with (a) 30 satellites (two minutes of link interruption occurs as presented in Table \ref{tab:OrbitSelection30}), (b) 40 satellites (continuous link is achieved as exhibited in Table \ref{tab:OrbitSelection40}).}
\label{fig:3}
\end{figure*}

\begin{algorithm}[!t]
\setstretch{1.35}
	\caption{Satellite Selection Algorithm}\label{alg:SatSel}
\begin{algorithmic}
    \For {$n=1,\ldots,N$}
        \State ${z}[n]$ $=$ $\mathbf{Z}[n,1]$ ;
        \State $\psi[n]$ $=$ $\mathbf{\Psi}[n,1]$ ;
        \State $\mathbf{S}[n\,,\,:]$ $=$ [\,$z[n]$ \, $\psi[n]$\,] ;
        \If{$M>1$}  \Comment{More than one available satellites}
		  \For {$m=2,\ldots,M$}
                \If{$\mathbf{Z}[n,m]$ $<$ $z[n]$}
                    \State $z[n]$ $=$ $\mathbf{Z}[n\,,\,m]$ ;
                    \State $\psi[n]$ $=$ $\mathbf{\Psi}[n\,,\,m]$ ;
                    \State $\mathbf{S}[n\,,\,:]$ $=$ [\,$z[n]$ \, $\psi[n]$\,] ;
                \EndIf
            \EndFor	
        \EndIf
    \EndFor
\end{algorithmic} 
\end{algorithm}

\vspace{-0.25 cm}

\subsection{Satellite Selection Method}
\label{Sec:sysmod:SSM}

Until now, the received power, harvested power, and system efficiency metrics have been discussed for the tracking and fixed solar arrays. After the assessment of eight different orbital schemes, it is concluded that single, triple, and quadruple orbits with 40 LLO satellites can establish an uninterrupted laser-based WPT to the solar array attached to the lunar rover at LSP. These multi-orbiter schemes allow multiple connections between LLO satellites and the rover simultaneously. Hence, this enables us to apply a satellite selection algorithm based on the shortest LoS distance and, hence, the smaller AoI for providing the optimum efficiency. 

The satellite selection approach that is used in our proposed model is described in Algorithm \ref{alg:SatSel}. The scalar time index $n$ increases up to the scalar constant $N$, and the satellite index $m$ can reach $M$, which denotes the number of total available satellites that have LoS links with the solar array on the lunar rover at $n$. The matrices $\mathbf{Z}$ and $\mathbf{\Psi}$ store all LoS distances to the solar panel and the incidence angles, respectively, for the available $M$ satellites at each $n$. Each ${z}[n]$ is defined as the minimum LoS distance at each $n$ among $M$ satellites from the matrix $\mathbf{Z}$. The angle of incidence $\psi[n]$ is determined from the matching satellite that provides the minimum LoS distance ${z}[n]$ at each $n$. The $N\times2$ satellite selection matrix $\mathbf{S}$ stores both ${z}[n]$ and $\psi[n]$, which are used in Eq. (\ref{EQ:6}) and (\ref{EQ:12}).  
 
The single and selected satellite visibility intervals for 30 and 40 orbiters in quadruple orbit configurations are presented in Fig. \ref{fig:3}. According to Fig. \ref{subfig:3a}, there are 2 minutes of link interruption in the 30-satellite scheme since there is no connection between the receiver at LSP and any LLO satellite at 21:25 and 21:26 time samples with 1 minute duration. Interestingly, the very same figure proves the significant change in the deployment of the orbiters based on the order change in the satellite visibility intervals on the 21\textsuperscript{st} day, especially when the time intervals in Fig. \ref{subfig:3a} and Fig. \ref{subfig:3b} are compared. The consecutive handover between the LLO satellites, which have simultaneous connections with the solar panel, can be seen in Fig. \ref{subfig:3b} on the 1\textsuperscript{st} day; however, that does not guarantee equal selected satellite visibility intervals after the SSM is applied.

\label{Sec:sysmod}
\vspace{-0.4 cm}
\section{Results and Discussion}
\begin{table}[!t]
  \sisetup{group-minimum-digits = 4}
  \centering
  \caption{WPT Simulation Parameters}
  \label{table:harvestedpower}
  \begin{tabular}{lllS[table-format=5]ll} 
    \toprule
    \toprule
Transmit power ($P_I$)                                                     & 1 kW 
\\ 
Laser wavelength ($\lambda$)                                                 & 1064 nm \cite{donmezatp}                               \\ 
Laser PCE (${{\eta }_{t}}$)                                                      & 51\% \cite{donmezatp}                              \\ 
Solar cell PCE (${{\eta}_{r}}$)                                                                 & 68.9\% \cite{lopez2023lunar}                           \\ 
\begin{tabular}[c]{@{}l@{}}Laser telescope \\ aperture diameter ($d_l$) \end{tabular}     & 0.3 m 
\\ 
\begin{tabular}[c]{@{}l@{}}Circular solar \\ array diameter ($d_r$)\end{tabular} & 2 m        \\
\begin{tabular}[c]{@{}l@{}}Simulation duration \end{tabular} & 27.3 days        \\
\begin{tabular}[c]{@{}l@{}}Simulation sample time \end{tabular} & 1 minute        \\
   \bottomrule
    \bottomrule
\end{tabular}
\vspace{-0.5 cm}
\end{table} 

Heretofore, 8 different Cislunar constellations are assessed to be able to establish a continuous FSO-based WPT link between the solar array and at least one of the LLO satellites throughout a Moon revolution around the Earth. Then, the mathematical models of the time-varying harvested powers and overall system efficiencies for the tracking and fixed solar arrays are discussed. As there are multiple simultaneous satellites that can establish WPT links, the SSM is introduced along with its algorithm. The handover of the links after the application of the SSM is demonstrated in Fig. \ref{fig:3} for the quadruple orbit scheme with 30 and 40 satellites. Besides, Fig. \ref{subfig:3a} shows the interruptions in 2 indices out of the 273 indices presented in Table \ref{tab:OrbitSelection30}. Now, the simulation settings will be explained briefly, and then the corresponding results will be presented and discussed in detail.   
 
Simulation parameters for harvested power calculations are presented in Table \ref{table:harvestedpower}. The laser wavelength of 1064 nm is selected due to its advantages in power conversion efficiencies as well as its optimum performance. The transmitter telescope aperture radius of 15 cm is selected, whereas the circular solar array radius is chosen as 1 m. Regarding the simulation duration, an approximate full moon revolution of 27.3 days \cite{MoonEarthFactSheetNASA} is considered, which makes our work realistic and applicable in Cislunar space, as the results from a limited simulation duration, such as a day, cannot be applied straightforwardly for a long duration, such as a month.  
   
\begin{figure*}[!t]
\centering
\subfloat[]{
	\label{subfig:4a}
	\includegraphics[clip, scale=0.75]{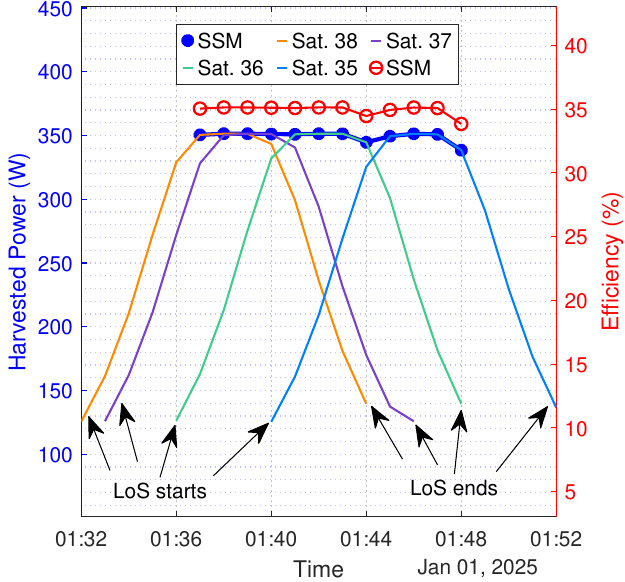}
	 }
\subfloat[]{
	\label{subfig:4b}
	\includegraphics[clip, scale=0.75]{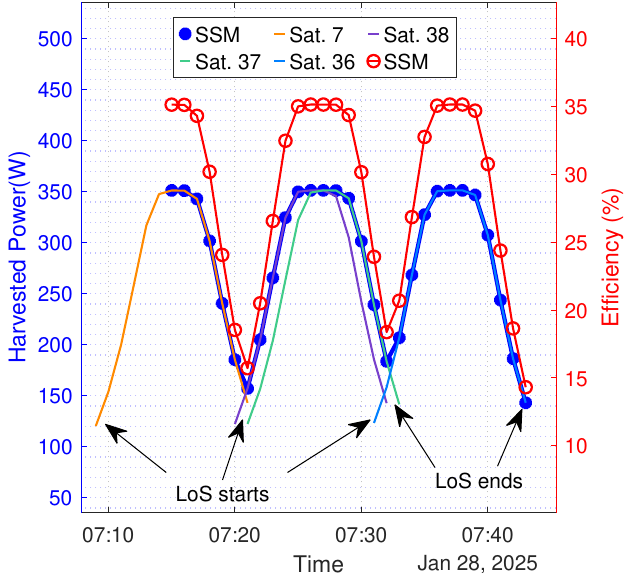}
	 }
\caption{The impact of the SSM over harvested power (blue) and overall system efficiency (red) for a tracking solar panel (a) day 1 (initial orbiter deployment and values of the mentioned metrics are maintained with SSM), (b) day 28 (initial orbiter deployment is not maintained thus SSM can only makes slight improvement).}
\label{fig:4}
\end{figure*}

In Fig. \ref{fig:4}, the SSM impacts over the harvested power and system efficiencies as a function of time are presented for the quadruple orbit with a 40-satellite scheme, which provides continuous WPT, for the tracking solar array case. The LoS link intervals for Satellites 7, 35, 36, 37, and 38 are presented in the figures. The harvested powers and system efficiencies of these individual satellites exhibit rollercoaster behavior as the visibility intervals are approximately 12 minutes. However, the satellite selection algorithm aims to extract the optimum values at each time index, and hence, the harvested powers (blue) and system efficiencies (red) are approximately 350 W and 35\%, respectively, on day 1 in Fig. \ref{subfig:4a}. On the other hand, a similar linear behavior cannot be seen in the last day in Fig. \ref{subfig:4b}, which proves that single-day simulation results cannot be carried straightforwardly for longer simulation durations. The harvested powers (blue) and system efficiencies (red) can drop to 157.2 W and 15.72\%, respectively, even after the SSM is applied. This disparity occurs because the initial deployment of the satellite constellation changes due to time-varying aformentioned external forces applied on the orbiters while the Moon both spins around itself and revolves around the Earth. Therefore, the initial equal distances between satellites on each orbit are not maintained thus, SSM is forced to make a selection out of one or two satellites rather than five as the time approaches 08:00 on Jan 28, 2025.

\begin{table}[!t]
\vspace{-0.15 cm}
\centering
\caption{Average Harvested Power Comparison Over 27.3 Days (A Moon Revolution)}
\label{Table:Comparison}
\begin{tabular}{|l|c|c|}
\hline
\multicolumn{1}{|c|}{\textbf{Configurations}}  &
  \textbf{\begin{tabular}[c]{@{}c@{}}Average \\ Harvested \\ Power (W)\end{tabular}} &
  \textbf{\begin{tabular}[c]{@{}c@{}}LoS Loss \\ Duration\\ (min.)\end{tabular}} \\ \hline
\begin{tabular}[c]{@{}l@{}}Single satellite with \\ tracking solar panel\end{tabular}           & 28.39  & 34,946 \\ \hline
\begin{tabular}[c]{@{}l@{}}SSM applied 30 satellites \\ with tracking solar panel\end{tabular}     & 323.26 & 273    \\ \hline
\begin{tabular}[c]{@{}l@{}}SSM applied 40 satellites \\ with tracking solar panel\end{tabular}      & 332.86 & 0      \\ \hline
\begin{tabular}[c]{@{}l@{}}Single satellite with \\ fixed solar panel\end{tabular}         & 9.68   & 34,946 \\ \hline
\begin{tabular}[c]{@{}l@{}}SSM applied 30 satellites \\ with fixed solar panel\end{tabular}  & 183.29 & 273    \\ \hline
\begin{tabular}[c]{@{}l@{}}SSM applied 40 satellites \\ with fixed solar panel\end{tabular}  & 204.43 & 0      \\ \hline
\end{tabular}
\vspace{-0.35 cm}
\end{table}

In Table \ref{Table:Comparison}, the $\overline{{P_{H_T}}}$ and $\overline{{P_{H_F}}}$ for a single satellite, 30 and 40 satellites in quadruple orbit scheme with SSM application during 27.3 days. The total number of time indices in which any LoS link is not achieved is 34,946 and 273 for a single and 30-satellite schemes, respectively, thus, there is no power reception for WPT. $\overline{{P_{H_T}}}$ for a single satellite, 30 and 40 satellites are 332.86~W, 323.96~W and 28.39 W, respectively. On the other hand, $\overline{{P_{H_F}}}$ for a single satellite, 30 and 40 satellites are 204.43~W, 183.29 W and 9.68 W, respectively. This decline occurs to the emerging $\psi$[n] as a result of the non-perpendicular reception of the laser rays by the fixed solar panel. 

\begin{figure*}[t!]
\centering
\includegraphics[width=0.9\textwidth]{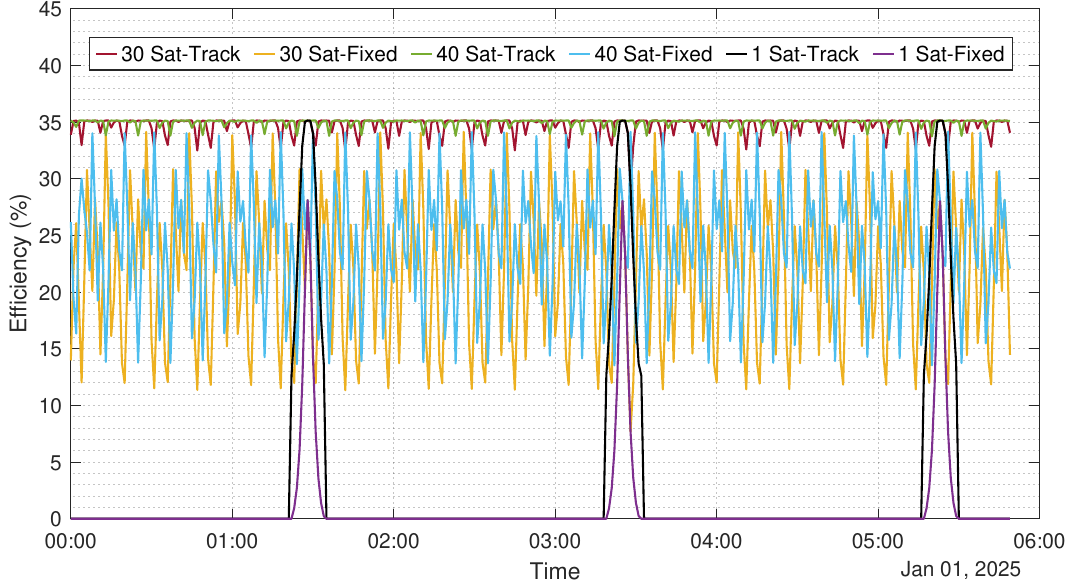}
\caption{Overall system efficiencies as a function of time for a single, 30 and 40 satellites schemes with SSM applied.}
\label{fig:5}
\end{figure*}

In Fig. \ref{fig:5}, the comparison of the time-varying overall system efficiencies ${\zeta_T}$[n] and ${\zeta_F}$[n] for a single satellite, 30-satellite, and 40-satellite are presented. Regarding the single satellite, there are only slight changes between the three cycles; the peak of ${\zeta_T}$[n] and ${\zeta_F}$[n] are 35.14\% and 28.11\%, whereas outside of the cycle time intervals, the received power and, hence, the efficiency becomes zero as there are LoS link losses. As satellite selection is applied to 30- and 40-satellite schemes, ${\zeta_T}$[n] do not drop to zero unless there is a LoS link loss. There are only slight fluctuations in ${\zeta_T}$[n] between maximum and minimum ${\zeta_T}$[n] for 30-satellite and 40-satellite schemes as 29.34--35.14\% and 33.74--35.14\%, respectively. However, there are considerable fluctuations between the extremes of ${\zeta_F}$[n] for 30-satellite and 40-satellite schemes as 7.67--34.14\% and 13.53--34.17\%, respectively. The peak ${\zeta_T}$[n] and ${\zeta_F}$[n] seem closer, yet $\psi$[n] is still larger than zero.



\label{Sec:results}
\vspace{-0.5 cm}
\section{Conclusion}
In our work, we investigated laser-based continuous wireless power transmission between the circular solar panel on the lunar rover and the optimum low lunar orbit satellite in terms of the time-dependent harvested power and overall system efficiency for a time interval of 27.3 days, or a revolution of the Moon around the Earth. Many Cislunar constellations were assessed to be able to determine which ones achieve a continuous link between at least one orbiter and the solar array located at the lunar south pole. The systems tool kit high-precision orbit propagator, which considers third body gravitational forces, solar radiation pressure, and Moon radiation pressure, was utilized while designing 8 different constellations, which are single, double, triple, and quadruple orbits with 30 and 40 orbiters. It was found that 40-satellite configurations allow the achievement of continuous lunar WPT. Then, the mathematical models to compute collected power, harvested power, and overall system efficiency were discussed for the solar panels with and without tracking ability. Regarding the fixed panel calculations, “the cosine effect” was incorporated. In addition, the satellite selection algorithm, which aims to activate only the optimum satellite offering the highest system efficiency among many others that also achieve simultaneous LoS link with the receiver, was introduced. 

The simulations in our work lasted 27.3 days, and they showed us an intriguing consequence in which the initial satellite deployment was not maintained and even changed significantly, as presented in both Fig. \ref{fig:3} and \ref{fig:4}. This approach made our computations of the average system efficiencies realistic, as not any short-duration simulation results were carried forward for approximation. As a result, the achieved average system efficiencies of a single, 30-satellite, and 40-satellite schemes are 2.84\%, 32.33\%, and 33.29\%, respectively, for the tracking panel and 0.97\%, 18.33\%, and 20.44\%, respectively, for the fixed solar panel case. 
\label{Sec:conc}
\section*{Acknowledgment}
This work was supported in part by the Tier-1 Canada Research Chair program.

\bibliographystyle{IEEEtran}
\bibliography{References}
\vspace{-0.25 cm}

\begin{IEEEbiography}
[{\includegraphics[width=1in,height=1.25in,clip,keepaspectratio]{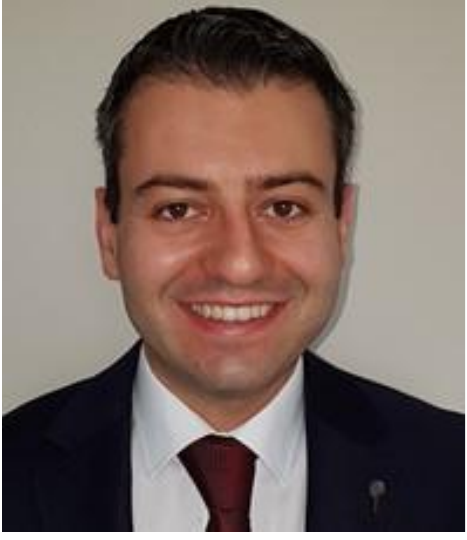}}] {Barış Dönmez}{\space}(Graduate Student Member, IEEE) received the B.Sc. and M.Sc. degrees (with high honors) in electrical and electronics engineering, in 2009 and 2022, respectively, from FMV Işık University, Istanbul, Turkiye. He is currently pursuing his Ph.D. degree in electrical engineering at Polytechnique Montréal, Montréal, QC, Canada. His research interests include communication and energy harvesting systems in space networks.
\end{IEEEbiography}

\begin{IEEEbiography}
[{\includegraphics[width=1in,height=1.25in,clip,keepaspectratio]{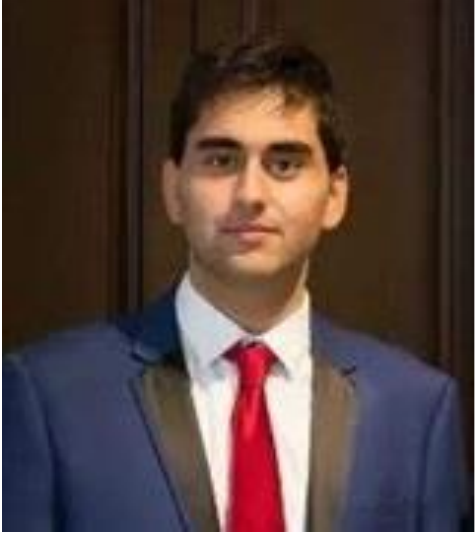}}] 
{Yanni Jiwan-Mercier}{\space} received his B. Eng. in physics engineering in 2020 and his M. Eng. in aerospace engineering in 2023, from École Polytechnique Montréal, Québec, Canada. He is currently doing his PH.D. degree in electrical engineering at Polytechnique Montréal on the subject of power transfer on the Moon. 
\end{IEEEbiography}

\begin{IEEEbiography}
[{\includegraphics[width=1in,height=1.25in,clip,keepaspectratio]{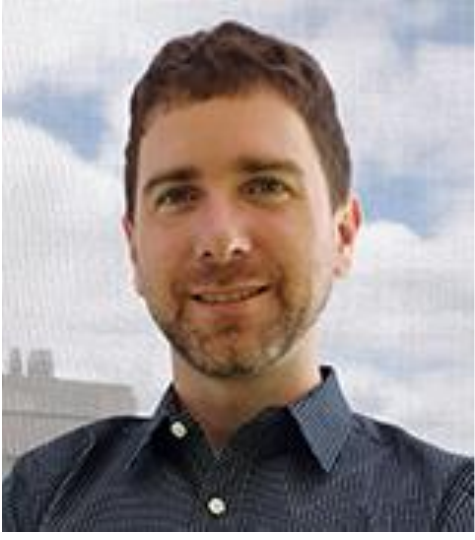}}] 
{Sébastien~Loranger}{\space}(Member, IEEE) is a Assistant Professor at Polytechnique Montreal, Montréal, Qc, Canada. He received Ph.D. degree in engineering physics in Polytechnique Montreal in 2018 and completed a post-doctoral fellowship at Max Planck Institute for the Science of Light, Erlangen, Germany in 2020. His research focus is in photonics and optoelectronic components for space.  
\end{IEEEbiography}

\begin{IEEEbiography}
[{\includegraphics[width=1in,height=1.25in,clip,keepaspectratio]{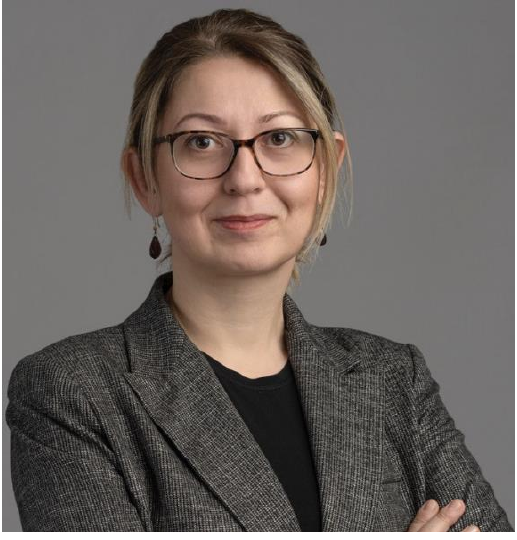}}] {Güneş Karabulut Kurt}{\space}(Senior Member, IEEE) is a Canada Research Chair (Tier 1) in New Frontiers in Space Communications and a Professor at Polytechnique Montréal, Montréal, QC, Canada. She is the Director of the Poly-Grames Research Center, and is co-founder and Director of Education and Training of ASTROLITH, Transdisciplinary Research Unit of Space Resource and Infrastructure Engineering at Polytechnique Montréal. She is also an adjunct research professor at Carleton University, Canada. Gunes received the B.S. degree with high honors in electronics and electrical engineering from Bogazici University, Istanbul, Turkiye, in 2000 and the M.A.Sc. and the Ph.D. degrees in electrical engineering from the University of Ottawa, ON, Canada, in 2002 and 2006, respectively. She worked in different technology companies in Canada and Turkiye between 2005 and 2010. From 2010 to 2021, she was a professor at Istanbul Technical University. Gunes is a Marie Curie Fellow and has received the Turkish Academy of Sciences Outstanding Young Scientist (TÜBA-GEBIP) Award in 2019. 
\end{IEEEbiography}
\label{Sec:bio}

\end{document}